\documentclass[11pt]{article} 
\usepackage{amsfonts,amscd,amsmath,graphicx}

\def\NP#1#2{ Nucl.Phys. B#1 (#2)}

\def\MPL#1#2{ Mod.Phys.Lett.A#1 (#2)} 
\def\PRL#1#2{ Phys.Rev.Lett. #1 (#2)} 
\def\PR#1#2{Phys.Rev. D#1 (#2)}

\def\ap{ \alpha^{\prime}} 

\def\pd{\partial}

\newcommand{\ep}{\text e}
\newcommand{\oh}{\frac{1}{2}}
\def\pp{\varphi{\prime}}
\def\3{\Phi^3}

\def\proc{AB\rightarrow CD}
\def\pp{p_{\perp}}
\def\ha{\hat\alpha}

\title{Scaling laws in hadronic processes and string theory}
\author{Oleg Andreev\thanks{e-mail:  andreev@physik.hu-berlin.de}
\thanks{Permanent address: Landau Institute, Moscow, Russia}
\\ \\
Humboldt--Universit\"at zu Berlin, Institut f\"ur Physik\\
Invalidenstra\ss e 110, D-10115 Berlin, Germany}
\date{}
\begin{document} 
\vspace{-8cm} 
\maketitle 
\begin{abstract} 
We propose a possible scheme for getting the known QCD scaling laws within string theory. In particular, we 
consider amplitudes for exclusive scattering of hadrons at large momentum transfer, hadronic form factors 
and distribution functions. 
\end{abstract}

\vspace{-10cm}
\begin{flushright}
hep-th/0209256     \\
HU Berlin-EP-02/42
\end{flushright}
\vspace{9cm}


\section{Introduction}
\renewcommand{\theequation}{1.\arabic{equation}}
\setcounter{equation}{0}
As it is well-known, string theory (dual-resonance model) was originally invented to describe the 
physics of hadrons \cite{dual}. However, in spite of much effort this idealized theory of hadrons failed and finally 
it was replaced by QCD. 

New ideas came in 1997-99; with Polyakov proposal for string theory whose tension is running, 
Maldacena conjecture of the AdS/CFT correspondence, and the Randall-Sundrum proposal for the hierarchy 
problem \cite{pol,mald,rs}. The key feature of all these is the warped geometry in spacetime, i.e., 
spacetime metric is no longer Minkowskian rather the normalization of the four-dimensional Minkowski 
metric is a function of other coordinates. In the simplest case, for example, the worldsheet 
action (its bosonic part) for the theory with the running tension  has the form
\begin{equation}\label{pol}
S=\frac{1}{4\pi }\int d^2z\,
\biggl(\pd\varphi\bar\pd\varphi +a^2(\varphi)\,\pd X\cdot\bar\pd X \biggr)
\quad,
\end{equation}
where $a^2(\varphi)$ is the running string tension. A natural requirement is that 
$a(\varphi)\sim\ep^\varphi$ as $\varphi\rightarrow\infty$. In other words, spacetime behaves for large 
$\varphi$ as $\text{AdS}_5$. The field $\varphi$ was called in \cite{pol} as the Liouville field.

Using these ideas, Polchinski and Strassler recently initiated a new attempt to describe the physics of 
hadrons in the framework of string theory \cite{ps} \footnote{See also \cite{gid,psu,braga}.}. They proposed  
a scheme of evaluating high-energy scattering 
amplitudes of glueballs in terms of vertex operators and found that the amplitudes fall as powers of 
momentum. This is the desired result which was found in the physics 
of hadrons long time ago \cite{mat}. To be more precise, the amplitude for exclusive scattering of $m+2$ glueballs
 is given by \cite{ps}
\begin{equation}\label{ps-amp}
{\cal M}(2\rightarrow m)\sim 
\frac{\left(gN_c\right)^{(n-2)/4}}{N_c^m\Lambda^{m-2}}
\biggl(\frac{\Lambda}{p}\biggr)^{n-4}
\quad,
\end{equation}
where $p$ is a large momentum scale, $g$ is the string coupling constant which is a square of the 
gauge coupling, $N_c$ is a number of colors, and $\Lambda$ is a scale by the lightest glueball. 
$n$ denotes a total number of constituents in the glueballs. 

In fact, it was known already in the seventies that dimensional analysis and some simple 
assumptions (dimensional counting rules) immediately lead to the correct scaling in exclusive hadronic 
scattering at large momentum transfer \footnote{See, e.g., \cite{b-far}.} 
\begin{equation}\label{amp-sc}
{\cal M}(2\rightarrow m)\sim p^{-n+4}
\quad.
\end{equation}
Indeed, if the total number of constituents in the hadrons is $n$ and one-particle 
state $\vert p\rangle$ is normalized 
in such a way to be dimension of $\text{[length]}$, then the amplitude has 
dimension $\text{[length]}^{n-4}$. If, moreover, at large momentum transfer, $p$ is the only length 
scale, then it immediately follows the wanted scaling. Modulo soft violations (logarithms) this scaling is 
in rather good agreement with experimental results.

It is now clear that the result \eqref{ps-amp} doesn't completely agree with this dimensional analysis of exclusive 
processes because it contains two dimensionfull parameters. It seems more relevant for inclusive 
processes where the second parameter is normally interpreted as the missing mass. In this case 
$n$ represents a subset of  constituents which participate in scattering; others remains ``spectators''. 
Moreover, in section 2 we will also see that QCD analysis provides a dependence on the coupling 
constant which differs from the one of Eq.\eqref{ps-amp}. 

The purpose of the present paper is twofold. The first is to propose a possible scheme for getting the scaling 
within string theory which resolves these difficulties. This scheme can be considered as a refinement 
of \cite{ps}. Like old matrix models (2d gravity coupled to conformal matter), where the scaling is obtained via 
a zero mode of the Liouville field \cite{ddk}, here we also get the scaling by a zero mode. As early mentioned, 
the warped geometry provides a natural candidate for a role of the Liouville field.  Note that this 
significantly simplifies the analysis as a knowledge of the whole dependence of vertex operators on 
the Liouville field is not required. As to the zero mode dependence, it is provided by the corresponding 
Laplace equation in spacetime. Thus our derivation of the scaling seems quite universal. The second is 
to apply this scheme to hadronic form factors and distribution functions for deep inelastic scattering. 
There are some special limits where these objects can easily be analyzed \cite{bjor, dy}. So we are bound 
to learn something if we succeed.

It is also worth mentioning that while this paper was being written, Polchinski and Strassler came up with 
a new paper \cite{ps3} \footnote{See also \cite{ps2}.} which has some overlap with what we describe 
in sections 2 and 3.

The outline of the paper is as follows. We start in section 2 by recalling some basic facts on QCD analysis of 
large-momentum-transfer processes. Here we focus on hard subprocesses for hadronic scattering and form 
factors. In Section 3 we present the scheme for getting the scaling behavior within string theory. 
We find complete agreement with the Born approximation in QCD. Section 4 contains our conclusions 
and a list of open problems. 

\section{Large-momentum-transfer processes in QCD}
\renewcommand{\theequation}{2.\arabic{equation}}
\setcounter{equation}{0}
In this section we will briefly recall some basic facts on QCD analysis of 
large-momentum-transfer exclusive processes \footnote{For further discussion, backgrounds 
and experimental data see, e.g., \cite{rev,rev2}.}. We will not consider spin dependent effects in 
what follows.
\subsection{Fixed-angle hadronic scattering}
In QCD analysis of a hadronic process $\proc$ the fixed-angle scattering 
amplitude related to hard subprocesses is given by the amplitude $T_H$ for the scattering of 
hadronic constituents, integrated over possible constituent momenta adding up to the hadron 
momenta. Explicitly, 
\begin{equation}\label{amp}
{\cal M}(\proc )=
\prod_{i=A,\dots,D}\int_0^1[dx^i]\,
\Phi_D^*(x^D,\pp^2)\Phi_C^*(x^C,\pp^2)
T_H(x^A,\dots,x^D,\pp^2)
\Phi_B(x^B,\pp^2)\Phi_A(x^A,\pp^2)
\,\,,
\end{equation}
where $x^i=\{x_1^i,\dots,x_{n_i}^i\}$, 
$[dx^i]=dx^i_1\dots dx^i_{n_i}\delta\left(1-\sum_{k=1}^{n_i}x^i_k\right)$, 
$\pp^2=tu/s$, and $n_i$ is a minimal number of constituents (valence quarks) in 
the $i$-th hadron \footnote{Note that adding more constituents (nonvalence quarks) is 
unimportant to leading order in $\pp^2$ at $\pp^2\rightarrow\infty$. So, $n_i$ every time 
means a minimal number.}. The momentum transfer between hadronic constituents occurs 
via the hard scattering amplitude $T_H$ which, to leading order in the coupling constant, is 
given by the sum of all Born diagrams with hadrons replaced by their constituents. A typical 
Born diagram looks like that in Fig.1. The amplitude $\Phi_i$ is a probability amplitude for 
finding constituents with fractions of longitudinal momenta $x^i_k$ in the $i$-th hadron. 
\begin{figure}[ht]
\begin{center}
 \includegraphics{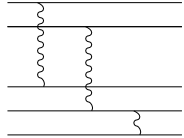}
 \caption{A typical Born diagram for meson-baryon scattering.  }
 \end{center}
\end{figure}

For $s\rightarrow\infty$ at $s/t$ fixed, $T_H$ falls with increasing $s$ as \footnote{Note that 
in the center of mass frame, $t=-s\sin^2\frac{\theta}{2}$, $u=-s\cos^2\frac{\theta}{2}$, and 
$\pp^2=\frac{1}{4}s\sin^2\theta$. So, $\pp^2\sim s$ at fixed angle.}
\begin{equation}\label{amp1}
T_H(x^A,\dots,x^D, \pp^2)=\biggl(\frac{e^2}{4\pi s}\biggr)^{\frac{n}{2}-2}
f(s/t,x^A,\dots,x^D)
\Bigl(1+O(e^2)\Bigr)
\quad,
\end{equation}
where $n=n_A+\dots +n_D$. Since the probability amplitudes in the Born approximation behave as
\begin{equation}\label{amp2}
\Phi_i(x^i, \pp^2)=\phi_i(x^i)
\quad,
\end{equation}
the amplitude ${\cal M}(\proc )$ takes the form
\begin{equation}\label{amp3}
{\cal M}(\proc )\sim \biggl(\frac{e^2}{4\pi s}\biggr)^{\frac{n}{2}-2}
\quad.
\end{equation}

At this point a comment is in order. It is well-known that radiative corrections in QCD typically 
contain logarithms that violates the scaling. To next order (one-loop approximation), they are included  
by replacing $e^2/4\pi\rightarrow\alpha_s(\pp^2)$, $\Phi_i(x^i, \pp^2)\rightarrow
\phi_i(x^i)\bigl(\ln\pp^2/\Lambda^2_{QCD}\bigr)^{-\gamma_i}$, where $\Lambda_{QCD}$ is 
the QCD parameter and $\gamma_i$ is some constant. Note that the amplitude can then be rewritten as 
\begin{equation}\label{amp4}
{\cal M}(\proc )\sim 
\bigl[\alpha_s(\pp^2)\bigr]^{\frac{n}{2}-2+\sum_i\gamma_i}\,\pp^{-n+4}
\quad.
\end{equation}
\subsection{Electromagnetic form factors and structure functions}

The electromagnetic  form factor of a hadron is given by a matrix element of the electromagnetic 
current between two hadronic states $\langle p+q\vert J^a(0)\vert p\rangle$. The form factors 
are most easily analyzed by using the two invariants $q^2=-Q^2<0$, $\nu_B=p\cdot q$ and then 
taking the Bjorken limit where $Q^2$ and $\nu_B$ both go to infinity with the ratio, $x_B=
Q^2/2\nu_B$ fixed \cite{bjor}. $x_B$ is known as the Bjorken variable. Note that  for elastic scattering 
$x_B=1$. Using Lorentz covariance and gauge invariance, the matrix element can be 
parameterized in terms of a scalar function (form factor) $F(Q^2)$ as 
\begin{equation}\label{form}
\langle p+q\vert J^\mu(0)\vert p\rangle=(2p^\mu+q^\mu)F(Q^2)
\quad, 
\end{equation}
where $J^\mu(0)=\int d^4k\,J^\mu(k)$.

To leading order in the coupling constant, the form factor in QCD takes the form 
\begin{equation}\label{form1}
F(Q^2)=
\int_0^1[dx][dy]
\,\Phi^*(x,Q^2)\, 
T_B(x,y,Q^2)\,
\Phi(y,Q^2)
\quad ,
\end{equation}
where $x=\{x_1,\dots,x_n\}$, 
$[dx]=dx_1\dots dx_n\delta\left(1-\sum_{k=1}^n x_k\right)$, and $n$ is a number of 
constituents (valence quarks) in the hadron \footnote{The variable $y$ and the integration 
measure $[dy]$ are defined in the same way.}. The amplitude $\Phi$ is defined in the same 
way as in the previous subsection while $T_B$ is now given by the sum of all Born diagrams for 
the hadron constituents to scatter with the photon producing the constituents in the final state. A typical 
Born diagram now looks like that in Fig.2.
\begin{figure}[ht]
\begin{center}
 \includegraphics{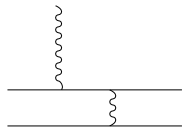}
 \caption{A typical Born diagram for the meson form factor.}
 \end{center}
\end{figure}

At large $Q^2$ the Born diagrams give 
\begin{equation}\label{form2}
T_B(x,y,Q^2)=\biggl(\frac{e^2}{4\pi Q^2}\biggr)^{n-1}
t_B(x,y)
\end{equation}
and thus, reasoning as in the previous subsection, the asymptotic behavior of the form factor 
is given by
\begin{equation}\label{form3}
F(Q^2)\sim
\biggl(\frac{e^2}{4\pi Q^2}\biggr)^{n-1}
\quad,
\end{equation}
 
A few noteworthy facts are the following. Dimensional analysis and the assumptions 
of section 1 which lead to the scaling law for the amplitudes can also be applied to the form factors. This 
immediately gives the desired scaling law (Bjorken scaling) $F(Q^2)\sim Q^{-2n+2}$. One of the most 
interesting applications of QCD was the prediction of slow violations of Bjorken scaling by soft 
$\ln Q^2$'s. To one-loop approximation, the logarithms are included  in the same way as in subsection 2.1: 
by replacing $e^2/4\pi\rightarrow\alpha_s(\pp^2)$, $\Phi_i(x^i, \pp^2)\rightarrow
\phi_i(x^i)\bigl(\ln\pp^2/\Lambda^2_{QCD}\bigr)^{-\gamma_i}$. Note that the form factor can 
also be rewritten as
\begin{equation}\label{form4}
F(Q^2)\sim
\bigl[\alpha_s(\pp^2)\bigr]^{n-1+2\gamma}
\left(Q^2\right)^{-n+1}
\quad.
\end{equation}

Let us conclude the discussion of QCD by briefly reviewing the hadronic structure functions. These are defined 
via the hadronic tensor
\begin{equation}\label{w}
W^{\mu\nu}(Q^2,\nu_B)=
\frac{1}{4\pi}\int d^4\xi\,\ep^{iq\cdot\xi}\,\langle p\vert \,J^\mu(\xi)\,J^\nu(0)\,\vert p\rangle
\end{equation}
as \footnote{Since we don't consider spin effects, we omit an antisymmetric part of $W^{\mu\nu}$ which provides 
two other structure functions: $g_1$ and $g_2$.}
\begin{equation}\label{w1}
W^{\mu\nu}(Q^2,\nu_B)=\biggl(-\eta^{\mu\nu}+\frac{q^\mu q^\nu}{q^2}\biggr)F_1(Q^2,\nu_B)+
\frac{1}{\nu_B}
\biggl(p^\mu-\frac{\nu_B}{q^2}q^\mu\biggr)
\biggl(p^\mu-\frac{\nu_B}{q^2}q^\nu\biggr)
F_2(Q^2,\nu_B)
\quad.
\end{equation}
Note that the functions $F_1$ and $F_2$ are related to others in common use by $W_1=F_1$ and 
$W_2=\frac{M^2}{\nu_B}F_2$, where $M$ is a hadronic mass.

The physical meaning of $F_2$ is that its $x_B$ dependence probes the longitudinal momentum distribution of 
the hadron constituents as viewed in the infinite momentum frame of the hadron. In particular, it is expressed in 
terms of the distribution functions $G_i$ as \footnote{$G_i(x_B)$ is defined as a probability of finding the $i$-th 
constituent in the hadron with fractional longitudinal momentum $x_B$ (in the infinite momentum frame of the 
hadron).}  
\begin{equation}\label{w3}
F_2(x_B)=x_B\sum_{i=1}^n\lambda^2_i\,G_i(x_B) 
\quad,
\end{equation}
where $\lambda_i$ is a charge of the $i$-th constituent. 

The structure functions are not known completely because they are in general beyond the tools of perturbative 
theory. However, some asymptotics are avaliable. In particular, the distribution functions 
which become functions (modulo soft logarithms) only of $x_B$ in the Bjorken limit,  behave as \cite{dy}
\begin{equation}\label{w4}
G_i(x_B)\sim\left(1-x_B\right)^{2n-3}
\end{equation}
near the threshold $x_B=1$. As before, $n$ means a total number of constituents in the hadron. It is worth 
mentioning that this asymptotic behavior can also be determined via a convolution equation for $G_i$ \cite{bb}.

\section{Scaling laws via string theory}
\renewcommand{\theequation}{3.\arabic{equation}}
\setcounter{equation}{0}
The aim of this section is to show how the Born approximation for hadronic amplitudes and 
form factors can be easily obtained in the framework of string theory.
\subsection{String theory settings}
According to our discussion of section 1, the metric asymptotically behaves as $\text{AdS}_5$. Since we are 
interested in the scaling rather than its violation, it is natural to use this metric. So, as in \cite{ps} 
we begin with string theory on the product of $\text{AdS}_5$ with a five-dimensional transverse space K. 
A spacetime metric is then
\begin{equation}\label{met}
ds^2=\frac{r^2}{R^2}\eta_{\mu\nu}dX^\mu dX^\nu+\frac{R^2}{r^2}dr^2+
R^2ds^2_{\text K}
\quad,
\end{equation}
where $R$ is a radius of $\text{AdS}_5$, $\eta$ is a four-dimensional Minkowski metric. We assume that 
K doesn't provide any dimensionfull parameter except $R$. Moreover, $ds^2_{\text K}$ doesn't depend 
on  $R$.

Before continuing our discussion of string theory settings, let us pause here to stress an important point. 
Although we use some technique inspired by the AdS/CFT conjecture, we don't strictly follow this 
conjecture. The point is that we are interested in the physical processes where perturbation theory is applicable 
rather than the strong coupling regime. So, we postulate the following relation
\begin{equation}\label{id}
g=\frac{e^2}{4\pi}=\frac{R^2}{\ap}
\end{equation}
between the closed string coupling constant $g$, the gauge coupling constant $e$, and the parameters $R$, 
$\ap$. This relation is not obviously what is imposed by the AdS/CFT correspondence. It becomes 
the latter (modulo a numerical factor) by replacing $e^2\rightarrow e,\,g\rightarrow\sqrt{g}$. 
It is worth mentioning that some examples where AdS/CFT results look like QCD ones after the above 
replacement are already known in the literature (see, e.g., \cite{malda,gkt}). 

In the first-quantized string theory one first introduces a free field action on a worldsheet, then defines physical 
vertex operators. Finally, scattering amplitudes in spacetime are defined as expectation values of the vertex 
operators. In general, it is unknown how to 
implement this program in the case of curved background like AdS. However, we think that the problem of 
interest doesn't require knowledge of the full string theory on AdS. It should be a simple stringy analog 
of the dimensional counting rules which results in the scaling laws. Our idea is to relate the scaling to a 
zero mode of $r$ as it is usually done in the context of 2d gravity where the scaling is due to a zero mode of 
the Liouville field. Then, all what we need is the dependence of vertex operators on this zero mode. The 
latter can be found from the Laplace equation on $\text{AdS}_5\times\text{K}$. In fact, our scenario 
means that non-zero modes of the transverse fields as well as $r$ are not of primary importance for 
the scaling. To leading order, they contribute a numerical factor. Alternatively, one can say that fluctuations 
of the transverse fields as well as $r$ are slow as it was done in \cite{ps}.

Under this assumption, it is straightforward to write down a part of the worldsheet action for the remaining
non-zero modes which is most appropriate for our purposes \footnote {Note that this is a conformal invariant 
action because $r$ doesn't depend on $z$.}
\begin{equation}\label{ac}
S=\frac{1}{4\pi }\int d^2z\,
\biggl(\frac{1}{\ha}\pd X\cdot\bar\pd X +\psi\cdot\bar\pd\psi+\bar\psi\cdot\pd\bar\psi\biggr)
\quad,
\end{equation}
where $\ha=\ap\frac{R^2}{r^2}$ and the $\psi$'s have been rescaled as 
$\psi\rightarrow \frac{R}{r}\psi$. We use this form of the action for two reasons: (1) It allows us to 
use the known results for string amplitudes simply by replacing $\ap\rightarrow\ha$. (2) It represents 
a model theory which has the running tension in the sense of Polyakov (see Eq.\eqref{pol}). 

To evaluate the correlation functions of vertex operators one needs to define the path integral measure. 
First let us do so for the zero modes \footnote{We only consider the NS-NS sector, so there are no zero 
modes of the $\psi$'s.}. It is natural to take it in a covariant form $\sqrt{-g}\,d^{10}\xi$. However, for 
a reason which will be clear in a moment we need it to be dimensionless. So, we define the measure as
\begin{equation}\label{measure}
\frac{1}{\ap{}^2R^6}\sqrt{-g}\,d^{10}\xi=\frac{1}{\ap{}^2R^4}\,r^3dr\,d^4x\,d\Omega_K
\quad,
\end{equation}
where $d\Omega_K$ is an invariant measure on K.  

The non-zero modes are quantized in an ordinary way as it follows from their 
action \eqref{ac}.The only novelty is an appearance of $\ha$ instead of $\ap$. For example, in the case 
of spherical topology the propagators are given by 
\begin{equation}\label{prop}
\langle X^\mu(z,\bar z)X^\nu(z',\bar z')\rangle=-\ha\,\eta^{\mu\nu}\ln\vert z-z'\vert
\quad,\quad
\langle\psi^\mu(z)\psi^\nu(z')\rangle=\frac{\eta^{\mu\nu}}{z-z'} 
\quad.
\end{equation}

As mentioned earlier, we are interested in the scaling properties of hadron interaction involving transfer of 
large momenta where all masses are negligible. Therefore the most appropriate string vertex operators 
to try are the massless ones. In general, one computes such vertex operators in the supergravity 
approximation (zero mode approximation) by finding solutions of the corresponding Laplace equation 
on $\text{AdS}_5\times\text{K}$. A general solution looks like 
$V\sim f(\Omega)\varphi(r)\ep^{ip\cdot x}$, where $\varphi$ shows a power-law falloff as 
$r\rightarrow\infty$. As mentioned earlier, the metric is $\text{AdS}_5$ only for large $r$, so it is pointless to use 
the exact solution for $\varphi(r)$ as long as we use the $\text{AdS}_5$ metric. The only thing we really need is 
its power-law falloff $\varphi\sim r^{-n}$. It was suggested in \cite{ps} that if one interprets vertex operators 
as hadronic states, one thinks of the $n$'s as the numbers of hadron constituents. The function $f$ is a solution 
of the Laplace equation on K which is responsible for internal degrees of freedom. Since we don't take them 
into account, we will not pay attention to $f$ as well. 

Now let us extend the analysis to include the non-zero modes. In the approximation we work it can easily be done by 
using the standard expressions for the vertex operators with $\ap$ replaced by $\ha$. Putting all this together, 
we get 
\begin{equation}\label{ver}
V_{n,\, p}=f(\Omega)\,
r^{-n}\,\ep^{ip\cdot x}
\int d^2 z\, \varepsilon_{\mu\nu}V_i^\mu(p,z)\bar V_i^\nu(p,\bar z)
\quad,
\end{equation}
where  $i$ means the superghost charge, $\varepsilon$ is the polarization tensor. Note that we extract 
the zero-mode factor $\ep^{ip\cdot x}$ from the vertex operators $V_i$ and $\bar V_i$ here and below. In general, 
the integrand of \eqref{ver} is more involved because $V_i$ and $\bar V_i$ are dressed by operators 
constructed from non-zero modes of $r$ and the transverse fields.

Since we don't consider spin dependent effects let us specialize to the dilaton vertex. To fix its 
normalization, we first make a rescaling $X\rightarrow\sqrt{\ha}X$ to bring the integrand into 
a dimensionless form. Then we insert a factor $\ap{}^n$. Thus, the vertex takes the 
form 
\begin{equation}\label{ver-d}
V_{n,\, p}=f(\Omega)\,
\biggl(\frac{\ap}{r}\biggr)^n\,
\ep^{ip\cdot x}
\int d^2 z\, \varepsilon_{\mu\nu}^{dil}\,V_i^\mu(\hat p,z)\bar V_i^\nu(\hat p,\bar z)
\quad.
\end{equation}
We use $\hat p$ as a shorthand notation for $\sqrt{\ha}\,p$. It is evident that such a vertex has 
dimension $[\text{length}]^n$ exactly as needed for the normalization of $n$-particle state. 

Now that we have the vertex operators for hadronic states, we can focus on the next object of interest, 
the vertex operator for the electromagnetic current. Since the current is conserved, it obeys
\begin{equation}\label{cur}
q\cdot J(q)=0
\quad.
\end{equation}
A natural realization which satisfies such a condition can easily be found in a picture where a string 
worldsheet admits boundaries \footnote{To our knowledge, this is the simplest way of introducing electromagnetic 
currents in string theory.}. In this case, we have
\begin{equation}\label{current}
J^\mu(q)\sim \ep^{iq\cdot x}
\oint_{C} dz\, V_0^\mu(\hat q, z)
\quad,
\end{equation}
where $V_0^\mu=\Bigl(\pd X^\mu+
\frac{i}{2}(\hat q\cdot\psi )\psi^\mu\Bigr)\ep^{i\hat q\cdot X}$, $\hat q=\sqrt{\ha}\,q$. $C$ denotes a 
worldsheet boundary.

By analogy with the vertex operators of hadrons, we insert the factor $f(\Omega)\Bigl(\frac{\ap}{r}\Bigr)^n$. Since 
the current has the dimensionality of $[\text{length}]$, we set $n=1$. Thus, a final form of $J^\mu(q)$ is given by 
\begin{equation}\label{current2}
J^\mu(q)=f(\Omega)\,
\frac{\ap}{r}\,
\ep^{ip\cdot x}
\oint_{C} dz\, V_0^\mu(\hat q, z)
\quad.
\end{equation}
\subsection{Evaluation of amplitudes}
The calculation of the scattering amplitude for a hadronic process $AB\rightarrow CD$ is mainly going 
along the lines of \cite{ps} adjusted to our settings. The amplitude is defined as the expectation value of 
the product of the vertex operators \eqref{ver-d}
\begin{equation}\label{s-amp}
\delta^{(4)}(p_A+\dots +p_D)\,{\cal M}(\proc )=\frac{1}{g^2}\,
\langle\,\prod_{i=A,\dots,D}\,g^{n_i}V_{n_i,p_i}\,\rangle
\quad.
\end{equation}
Here the string worldsheet is a sphere as is usual at tree level in closed string perturbation theory. Some factors of 
this expression require further explanation: (1) The over-all factor $g^{-2}$ comes from the sphere as it should 
be. (2) The factors $g^{n_i}$ are due to our normalization prescription. It differs from the standard one and will 
be discussed later.

The integration over non-zero modes doesn't require much of work at least in the case of spherical topology where 
the four-point dilaton amplitude ${\cal A}_4$ is well-known in the literature \cite{gs}. Aside from an irrelevant 
numerical factor, the  amplitude is then given by
\begin{equation}\label{s-amp1}
{\cal M}(\proc )=\frac{g^{n-2}}{\ap{}^2R^4}\int_0^\infty dr\,r^3\,
 \biggl(\frac{\ap}{r}\biggr)^n
{\cal A}_4(\ap R^2s/r^2,\ap R^2t/r^2,\ap R^2u/r^2)
\quad.
\end{equation}
Here the integral over $\Omega_K$ has not disappeared, but was included in the numerical factor. $n$ is 
the sum of the $n_i$'s. From the above expression it is evident that if we rescaled $r$ as 
$r\rightarrow\sqrt{\ap s}R\,r$, then we get
\begin{equation}\label{s-amp2}
{\cal M}(\proc )=F(\theta)\Bigl(\frac{g}{s}\Bigr)^{\frac{n}{2}-2}
\quad,
\end{equation}
where $F(\theta)$  is a function of the angle $\theta$ defined in the center of mass frame. This is the desired result, 
and it is identical to the result of QCD, Eq.\eqref{amp3}.

A couple of comments is in order: 
\newline (1) The use of the relation \eqref{id} is crucial for matching with the QCD result.
\newline (2) It may appear that the scaling is due to the zero modes only. That is not exactly true. 
The none-zero modes contribute into the function $F(\theta)$ which contains some important information. 
We will return to this issue in section 4.

\subsection{Evaluation of form factors and distribution functions}
By analogy with the amplitudes, we define the form factor as the expectation value of the product of the vertex 
operators given by \eqref{ver-d} and \eqref{current2}. Explicitly,
\begin{equation}\label{for}
\langle p+q\vert\,J^\mu(0)\,\vert p\rangle=
\int d^4k\,\,
\frac{1}{g}\langle\, g^nV_{n,p+q}\,\sqrt{g}J^\mu(k)\,g^nV_{n,p}\,\rangle
\quad.
\end{equation}
The worldsheet is now a disk (upper half plane). So we insert the over-all factor $g^{-1}$ as is usual in the case of the 
disk. Just as before, each closed string vertex carries a factor $g^{n}$. From this, it seems natural to 
accompany each open string vertex with $g^{n/2}$. If so, then $J^\mu$ is accompanied by $g^{1/2}$.

To evaluate the right hand side of Eq.\eqref{for}, it is convenient to use the worldsheet doubling 
trick (see, e.g.,\cite{hk} and references therein). After performing the integration over $x$ and setting the vertex 
operators at $(z_1,\bar z_1, z_2, z_3, \bar z_3)=(iy, -iy, t, i, -i)$, we find (modulo a numerical factor)
\begin{equation}\label{for1}
\langle p+q\vert\,J^\mu(0)\,\vert p\rangle=
\frac{g^{2n-\oh}}{\ap{}^2 R^4}\int_0^{+\infty} dr\,r^3\,
\biggl(\frac{\ap}{r}\biggr)^{2n+1} {\cal A}^\mu(\hat p, \hat q)
\quad,
\end{equation}
with
\begin{equation}\label{for2}
{\cal A}^\mu(\hat p, \hat q)=
\int^{+\infty}_{-\infty} dt \int_0^1 dy\,
\varepsilon^{dil}_{\eta\nu}\,\varepsilon^{dil}_{\lambda\sigma}\,
\langle\,
V_{-1}^\eta(iy,-\hat p-\hat q)\,V_{-1}^\nu(-iy,-\hat p-\hat q)\,
V_0^\mu(t, 2\hat q)\,V_0^\lambda(i, \hat p)\,V_0^\sigma(-i, \hat p)
\,\rangle
\quad.
\end{equation}
Here we again include the integral over $\Omega_K$ into an irrelevant numerical factor. 

To keep things as simple as possible, first we choose the infinite momentum frame for the 
hadron \footnote{Note that $Q^2=-q^2=(q^1)^2+(q^2)^2+O(1/P^2)$.}
\begin{equation}\label{imf}
p^\mu=\left(P+M^2/2P, 0, 0, P\right)
\quad,\quad
q^\mu=\left(\nu_B/P, q^1, q^2, 0\right)
\quad.
\end{equation}
Here $M^2$ is a mass of the hadron. Secondly, we specialize in a convenient current component $J^0$ or, 
equivalently, $J^3$. Then, it follows from Lorentz covariance that 
$ {\cal A}^0(\hat p, \hat q)=\sqrt{\ha}P{\cal A}(\ha Q^2)$. 

Using Eq.\eqref{form}, we find the following representation for the form factor
\begin{equation}\label{for3}
F(Q^2)=\frac{g^{2n-\oh}}{\ap{}^{\frac{5}{2}}R^3}\int_0^{+\infty} dr\,r^3\,
\biggl(\frac{\ap}{r}\biggr)^{2n+2} {\cal A}(\ap R^2Q^2/r^2)
\quad.
\end{equation}
The desired QCD result \eqref{form3} is obtained by rescaling $r\rightarrow\sqrt{\ap Q^2}Rr$ and  
using the relation \eqref{id}.

At this point, it is necessary to make a couple of remarks. 
\newline (1) Unlike the four-point dilaton amplitude ${\cal A}_4$ we used to
evaluate the hadronic amplitudes in the previous subsection, the correlator of the five vertex operators in 
Eq.\eqref{for2} is not well-defined in the following sense. As an object of 2d conformal field theory, 
$\int dt\,V^\mu_0(t, 2\hat q)$ is well-defined only at $q^2=0$ while for our purposes we need it 
at large $q^2$. In fact, this is the old standing problem of string theory: how to continue correlators of vertex operators 
defined on-shell (at special values of momenta) to off-shell (for arbitrary values of momenta). So far, there is no 
solution to this problem. In the problem of interest it means that ${\cal A}^0$ is in general 
ambiguous \footnote{Because of this, it seems pointless to give an explicit calculation of ${\cal A}^0$. It will 
suffer from ambiguity as any off-shell continuation.}. However, it is clear from the above discussion that 
the explicit form of ${\cal A}^0$ is not of principal importance for our purposes. So, our results seem rather universal 
and independent of a special way of going off-shell. We will return to this point in Section 4.
\newline(2) It is straightforward to evaluate the inelastic form factors by using the same technique. It is clear that 
the result has the same form as before with $n$ replaced to $(n_1+n_2)/2$. Here $n_i$ means a number of 
constituents in the $i$-th hadron.

Finally, let us discuss how the asymptotics \eqref{w4} for the distribution functions can be obtained in 
string theory. In fact, it was realized long years ago \cite{dy} that  this asymptotics is closely related to the asymptotic 
behavior of the form factors we have just considered. Thus, it seems natural to reproduce it too. 

To do so, we first choose a convenient infinite momentum frame defined by Eq.\eqref{imf}. Our next task is to 
evaluate a probability amplitude of finding the $i$-th constituent in the hadron with fractional longitudinal 
momentum $x_B$. If $V_{n,p}$ describes a hadronic state with $n$ 
constituents, then the best what we can use as an approximation to the hadronic state containing the $i$-th 
constituent with a specific momentum is $V_{1,p'}V_{n-1,p-p'}$. What is important to remark is that unlike $p$, 
all other momenta are not light-like. Thus, the corresponding vertex operators are off-shell. The probability 
amplitude is simply 
\begin{equation}\label{pa}
A_i\sim
\langle\,
V_{n,p}V_{1,p'}V_{n-1,p-p'}
\,\rangle
\quad.
\end{equation}
To compute the distribution function, we have to integrate $\vert A_i\vert^2$ over momenta of the constituents. 
In our approximation to the probability amplitude there is no integration over longitudinal momenta as $x_B$ is 
fixed. As to transverse momenta, it seems natural to parameterize them in terms of $q$ (see \eqref{imf}). The lovely 
thing about the threshold $x_B=1$ is that at leading order in $1-x_B$ we can take $q=p-p'$. Just as before, it is now 
easy to evaluate the scaling behavior of the amplitude 
\begin{equation}\label{pa2}
A_i\sim\frac{1}{(Q^2)^{n-2}}
\quad.
\end{equation}
Finally, the desired result is obtained after a simple estimation
\begin{equation}\label{pa3}
G_i(x_B)\sim\int_{\frac{M^2}{1-x_B}}^{+\infty}dQ^2\,\, A_i^2
\sim\left(1-x_B\right)^{2n-3}
\quad.
\end{equation}
At this point, it is worth mentioning that in approaching the threshold one must satisfy the 
inequality $Q^2(1-x_B)>M^2$ in order to stay in the Bjorken limiting region for $x_B$. This inequality 
provides the lower limit of integration. 

\section{Many open problems}
\renewcommand{\theequation}{4.\arabic{equation}}
\setcounter{equation}{0}

There is a large number of open problems associated with the circle of ideas explored in this paper. In 
this section we list a few.  

It would be interesting to understand in more detail how string theory reproduces the results of QCD in the 
Born approximation. The point is that in QCD the calculation of the hadronic scattering amplitudes  
involves a summation of a huge number of Born diagrams like the one presented in Fig.1. On the other hand, 
we saw in section 3 that in string theory the summation is automatically done and all information is encoded 
in the function $F(\theta)$.  Thus, this function may be considered as a generating function for Born diagrams. 
If so, it would significantly simplify ordinary QCD calculations. The problem is how to implement this explicitly. 
Unfortunately, our approximation is invalid for computing the exact form 
of $F(\theta)$. A possible way to deal with the problem is of course to involve the non-zero modes of $r$ 
and even the transverse fields. A price for this is a long standing problem: string theory on $\text{AdS}_5$. 
Though some information which is relevant for deep inelastic scattering has already been extracted from 
this theory (see, e.g., \cite{psu,gkp}), a complete solution is still missing. 

A related problem is understanding more clearly the stringy calculation of the form factors. Even without 
turning on the non-zero modes of $r$ and the transverse fields, it requires off-shell continuation. In principle, 
accounting for the non-zero modes might help with off-shell continuation. However, another interesting idea for 
doing so is to try the original Liouville mode as it comes from a conformal factor of the worldsheet metric \cite{am}. 
In addition, it would be interesting to compute the matrix element $\langle p\vert J^\mu(\xi)J^\nu(0)\vert p\rangle$ 
directly by using the vertex operators. 

It should be stressed that the string theory construction we are dealing with has an essential difference 
from the standard one. Usually each external leg of Feynman diagram corresponds to a vertex operator 
in the corresponding string correlator representing the amplitude. For example, one has at tree level in 
closed string perturbation theory
\begin{equation}\label{cou}
g^{m-2}\,\langle\, V_{n_1, p_1}\,\dots\,V_{n_m, p_m}\,\rangle
\quad,
\end{equation}
where each operator is accompanied by $g$. It is transparent from the diagrams of Figs.1-2 that in 
the problem of interest  we assigned a vertex operator also to a bunch of external legs. This is as it should be 
because hadrons are composite objects. As a consequence, our normalization 
prescription for the vertex operators is different of \eqref{cou}. It is clear that the 
standard prescription fails if it is blindly applied to recover the QCD results. To see what happened, consider our 
normalization in more detail. We begin with a modification of \eqref{cou} via replacing 
$g\,V_{n_i,p_i}\rightarrow g^{n_i}\,V_{n_i,p_i}$. This gives the over-all factor $g^{n-2}$, 
where $n=n_1+\dots+n_m$. However, this is not the whole story. The point is that the expectation value of the 
product of the vertex operators provides an additional factor $g^{-n/2}$. This effect is unknown in the case of 
Minkowski spacetime because it is due to the warped geometry. Thus, we end up with the desired answer. 
Modulo $g^{-2}$, the effect of the warped geometry is in fact the transformation of the closed string coupling 
constant into the open string coupling constant $g\rightarrow\sqrt{g}$. It would be interesting to see whether 
the warped geometry also transforms the open string coupling $e$ to $\sqrt{e}$. If so, then it might help to explain 
the known effect $e^2N_c\rightarrow\sqrt{e^2N_c}$ observed in the AdS/CFT 
calculations (see, e.g., \cite{malda,gkt}). 

Another interesting problem involves computing the quantum corrections. Our discussion here was entirely 
classical. At first glance, the QCD results of section 2 formally assume a slight modification of the Born approximation 
at one-loop level that on the string theory side can be implemented by just 
replacing $g^{n_i}\rightarrow g^{n_i+\gamma_i}$. But a real situation is much more involved. The point is that 
the coupling constant is now running. So, if we indeed wish to recover the QCD results, we need to provide a mechanism 
which makes the coupling running. One possible way is to deform the string background to get the desired running. 
How to implement this and what will happen remains to be seen. 

As we mentioned, we don't strictly follow the prescriptions of the AdS/CFT correspondence. So, we have postulated the 
relation \eqref{id}. Our motivation is that first of all we would try to describe the known results of QCD to see whether 
the ideas work or not. Note also that, one can exploit this relation to make a simple estimation how large is the 
internal compact space $K$ in terms of $\ap$
\begin{equation}\label{R}
R^2=0.1\,\ap
\quad.
\end{equation}
Here we used the fact that a typical value of the coupling constant obtained from deep inelastic scattering 
experiments is of order $0.1$. Next step would be to see what happens in the strong coupling regime where it is believed
that the AdS/CFT correspondence holds. At present we lack the solution of string theory on $AdS_5\times S^5$ or on 
its nonconformal deformations that could help us. However, let us nonetheless see what information about high-energy 
scattering can be found by using our approximation. Assuming as in \cite{ps} that at small $r$ the geometry given by 
\eqref{met} is truncated at $r=r_0$, the amplitude \eqref{s-amp1} then becomes
\begin{equation}\label{s}
{\cal M}(\proc )\sim\frac{g^{\frac{3}{4}n-2}}{N_c^{\frac{n}{4}}}
\biggl(\frac{1}{\sqrt s}\biggr)^{n-4}
\int_{r_0/\sqrt{\ap s}R}^\infty dr\,r^{3-n}\,
 {\cal A}_4\Bigl(\frac{1}{r^2}, -\frac{\sin^2\theta/2}{r^2},-\frac{\cos^2\theta/2}{r^2}\Bigr)
\quad,
\end{equation}
where we also rescaled $r$ as $r\rightarrow\sqrt{\ap s}Rr$. It is at least somewhat plausible in the hard 
scattering limit ($s\rightarrow\infty$) that $r_0/\sqrt{\ap s}R\ll 1$ \footnote{In the model described in 
\cite{ps}, $r_0=\Lambda R^2$, where $\Lambda$ is a scale by the lightest hadron. So, it means that 
$\Lambda g^{\frac{1}{4}}\ll\sqrt{s}$.}. If so, then the leading behavior of the amplitude is the same power-law 
falloff as in \cite{ps} \footnote{Note that the coupling dependence is different.}. It is of some interest to 
evaluate corrections to the scaling. To do so, first we note that at small $r$ the four-point amplitude ${\cal A}_4$ 
behaves as ${\cal A}_4\sim\ep^{-\frac{1}{r^2}f(\theta)}$. Next we estimate the correction as 
\begin{equation}\label{s2}
\frac{g^{\frac{3}{4}n-2}}{N_c^{\frac{n}{4}}}
\biggl(\frac{1}{\sqrt s}\biggr)^{n-4}
\int_0^{r_0/\sqrt{\ap s}R} dr\,r^{3-n}\,\ep^{-\frac{1}{r^2}f(\theta)}\sim
\ep^{-\frac{\ap R^2}{r_0^2}s f(\theta)}
\quad.
\end{equation}
Unlike in section 2, where the radiative QCD logarithms violate the scaling, there is now an exponential correction 
which violates the scaling already at tree level.

There is a list of other interesting issues as including spin, flavor, color; soft subprocesses; the evolution of the 
distribution functions, and many others which certainly deserve to be addressed. 

\paragraph{ Acknowledgments.} We have been benefited from discussions with H.B. Nielsen, J. Polchinski, 
and A.M. Polyakov. We also would like to thank M. Strassler for correspondence; H. Dorn and A.A. Tseytlin for 
comments and reading the manuscript. This  work  is supported in part by DFG under grant No. DO 447/3-1 
and the European Commission RTN Programme HPRN-CT-2000-00131. 


\end{document}